Everyday Uses of Music Listening and Music Technologies by Caregivers and People with Dementia: Survey and Focus Group Study

Dianna Vidas, Romina Carrasco, Ryan M. Kelly, Jenny Waycott, Jeanette Tamplin, Kate McMahon, Libby M. Flynn, Phoebe A. Stretton-Smith, Tanara Vieira Sousa, & Felicity A. Baker

**Abstract**

**Background:**

Music has long been identified as a non-pharmacological tool that can provide benefits for people with dementia, and there is considerable interest in designing technologies to support the use of music in dementia care. However, to ensure music technologies are appropriately designed for supporting caregivers and people living with dementia, there remains a need to better understand how music is currently used in everyday dementia care at home.

**Objective:**

This study aimed to understand how people living with dementia and their caregivers use music and music technologies in everyday caring, as well as the challenges they experience using music and technology.

**Methods:**

This study used a mixed methods design. A survey was completed by 77 caregivers and people with dementia to understand their use of music and technology. Subsequently, 18 survey respondents (12 family caregivers, 6 people living with dementia) participated in focus groups regarding their experiences of using music and technology in care. Interview transcripts were analysed using reflexive thematic analysis.

**Results:**

Most survey respondents (including both people living with dementia and their caregivers) said they used music often in their daily lives. Participants reported a range of technologies used for listening to music, such as CDs, radio, and streaming. Focus groups highlighted benefits and challenges of using music and music technologies in everyday care. Participants identified using music and music technologies to regulate mood, provide joy, facilitate social interaction and connection, encourage reminiscence, provide continuity before and after diagnosis, and to make caregiving easier. Challenges of using music technology in everyday caring included difficulties with staying up to date with evolving technology, and low self-efficacy for technology use expressed by people living with dementia.

**Conclusions:**

This study shows that people with a dementia diagnosis and their caregivers already use music and music technologies to support their everyday care needs. Results suggest opportunities to design technologies that enable easier access to music and to support people living with dementia with recreational and therapeutic music listening and music-based activities.

# Introduction

**Background**

Worldwide, over 55 million people are currently living with dementia, making dementia one of the leading causes of disability among older adults globally [1]. While many people living with dementia are older adults aged 65 and above, younger onset dementia can affect people of different ages. As well as difficulties with cognition, people living with dementia may experience a range of behavioural and psychological symptoms, such as agitation, anxiety, irritability, and depression [2]. These difficulties place considerable demands on family and other "informal" caregivers, who provide the majority of care for people with dementia living in the community [3]. Although informal caregivers frequently report that they need additional support and information [4], they are often reluctant or lack knowledge to fully utilise services [5], leaving many caregivers feeling burdened and experiencing depressive symptoms [6]. As a result, caregivers and people with dementia need accessible strategies to support their everyday care needs [7–10].

Music is a non-pharmacological tool that can provide numerous benefits for people with dementia [11–14]. *Therapeutic* music experiences, where music is used as an intervention by a trained therapist, have long been shown to assist in managing symptoms of dementia such as agitation and depression, and can support interpersonal connection between people with dementia and their caregivers [11,15–19]. While music therapy is a promising avenue for improving the lives of people with dementia and their caregivers, music therapy may not be readily accessible for all people with dementia, particularly those who are less able to participate in activities outside the home [20]. Caregiver training programs informed by music therapy, where caregivers are trained by music therapists in intentional uses of music, have received attention as a method of providing quality informal care [21].

*Recreational* music experiences, such as music listening, can also provide important everyday support for improving quality of life [22]. In the aged care setting, recreational music listening can provide people with dementia with enjoyment, relaxation, social connectedness, entertainment, and mood regulation [13,23]. Similar benefits of music use have been identified for older adults with dementia living at home, with music reducing agitation, improving cognition, and enhancing social wellbeing and connection [19,24]. Much of the research on this topic however has focused on identifying benefits gained from participating in organised music activities, such as community singing groups [18]. There has been less research investigating how family caregivers utilise music and music technologies in everyday home-based care [25]. One exception, a study by Elliott and colleagues [23], used a case study approach to examine the musical experiences of three people living with dementia and their caregivers, but did not examine specific music technologies. As such, there remains a need for further research to understand the role of music in everyday dementia care, and the role of technology in facilitating access to music for people with dementia and their caregivers.

While technology can support both therapeutic and recreational uses of music in dementia care, existing research has focused primarily on using technology to support recreational uses of music, primarily in formal care settings. Several technologies have been developed and trialled to support music listening and music-based activities for people living with dementia [26,27]. Researchers have designed and trialled technologies that aim to support music for dementia care by bringing enjoyment and social connection to people with dementia, through both music and sound [28,29]. In one example, Hodge et al. [30] created a virtual reality environment of a concert experience, tailored for the specific interests of one person living with dementia. Music-based technologies in the aged care context have been explored as a way to promote an enjoyable, active form of engagement with music that may facilitate connection and shared movement [31–33]. In another example, researchers created a musical interface that aimed to facilitate collaborative music-making to foster partnership between aged care residents and professional caregivers [29]. Interactive approaches to designing musical experiences for people living with dementia extend to mobile applications, which can facilitate collaborative music-making [34]. In addition, music therapists employ a range of technologies to engage their clients in communication, connection, and music-making [35], suggesting that music technologies have utility for recreational and therapeutic music experiences for people with dementia.

In this paper, we aim to understand the role of music and music technologies in the everyday lives of people with dementia and their family caregivers. Further, we consider the potential of music technologies both for recreation, and for managing behavioural and psychological symptoms of dementia. Music is now often accessed through technologies such as streaming services, but there may be barriers to using these technologies in the complex environment of family-based dementia care. To create new technologies that support the use of music for people living with dementia at home, we must understand how music technologies currently support people with dementia and their caregivers, and what challenges these groups face in using music with current technologies.

**Objectives**

This study used a mixed methods design with two parts. An initial survey with family caregivers and people with dementia aimed to investigate their use of music and technology. Survey participants were then invited to take part in the focus group study, which further discussed current practices with music and music technologies in dementia care. We aimed to understand how people with dementia and their family caregivers use music and music technologies as part of their everyday care practices. We further aimed to explore the perceived benefits of using music and music technologies, as well as the perceived challenges caregivers and people living with dementia face in accessing music through technology.

<div align="center">

**Survey Study**

**Method**

</div>

**Ethics approval**

All procedures were approved by the University Human Research Ethics Committee [removed for review].

**Participant Recruitment**

**Survey**

An anonymous online survey was distributed through Facebook advertisements and StepUp for Dementia Research (an online platform that connects volunteer research participants with dementia researchers in Australia) [36]. Survey respondents could also indicate their interest in participating in focus groups for the main study.

**Focus Groups**

Survey respondents who were willing to participate in the focus groups were sent an electronic plain language statement and consent form indicating they could be contacted via phone or email. Participants who provided this consent were then contacted by a member of the research team who invited them to take part in online focus group discussions. During this contact, the researcher provided an overview of the study and opportunity for participants to ask questions. Participants could attend the sessions alone or as a dyad (caregiver and person living with dementia attending together). All participants were offered an AUD$30 voucher for their time.

**Procedure**

**Survey**

A plain language statement and consent form were shown at the beginning of the survey. After providing consent, the survey asked participants for demographic information (age, gender, level of education). Next, participants were asked about how they currently use music in their daily life, whether music has been an important or meaningful part of their life in the past, their musical preferences, and technologies used to listen to music. Finally, participants were asked about their experience with common technologies, openness to using mobile applications to support daily activities, and ways that a music-based mobile application might be able to support daily activities. For some questions, caregivers answered both for themselves and the person they cared for, and people with dementia responded for themselves and their caregiver, if relevant. The survey took approximately 15 minutes to complete.

**Focus Groups**

Five focus group sessions were organised. All sessions were conducted via Zoom to enable the remote participation of people in different regions of Australia and to mitigate COVID-19 infection risk. Only one participant (a caregiver) attended the first focus group, as two other participants withdrew at the last minute due to unforeseen circumstances. The second and third focus groups were conducted with caregivers only. The fourth focus group was conducted with caregivers and one dyad (one person with dementia and their caregiver), and the fifth focus group was conducted with a mix of dyads and people living with dementia who attended the session independently (without a caregiver).

The focus groups followed a protocol divided into four phases. Each focus group started with an introduction (5–10 minutes) where participants and researchers introduced themselves. The second phase for caregivers involved discussing their role as a caregiver, experiences of using music and technology in care, types of music technologies currently used, and how they believed technology could support care. People living with dementia were similarly asked about their experiences with music and how they use technology to listen to music. The third phase involved discussion about how potential future music technologies could respond to their needs, leading into a wrap up and conclusion (phase four). The questions were developed by the research team to align with the broader goals of the project, which aimed to understand what caregivers and people with dementia wanted from a mobile music therapy informed application. All focus group sessions were video and audio recorded.

**Data Analysis**

Survey responses were analysed using descriptive statistics; we tabulated frequency data for responses to each question. Audio recordings of focus groups were transcribed by the researchers. We analysed the transcripts using reflexive thematic analysis [37,38]. The data were coded using an inductive approach. Researcher subjectivity is central to reflexive thematic analysis, which is typically led by a single coder [39]. The present analysis was led by the first author, who has training in the fields of human-computer interaction and music psychology, as well as experience conducting research with people living with dementia and their caregivers.

After familiarisation with the transcripts, the first author generated initial codes and developed initial themes, which included six themes grouped under the category of benefits of music and technology, and five themes grouped under the category of barriers to use. Initial themes were discussed among a subset of the authors with expertise in human-computer interaction. Initial themes were subsequently defined and named, and further refined, taking an iterative approach during writing. Further discussion with the remaining authors (with expertise in music therapy), and reflection on presentation of findings led to additional revision and finalisation of themes presented here.

**Results**

**Survey**

The survey was launched online in December 2020, and was open during the first quarter of 2021. There were 77 surveys from 13 people living with dementia and 64 caregivers. Of these, 9 people living with dementia and 24 caregivers (37.5%) consented to be contacted for the focus groups. Two people living with dementia and two caregivers also invited their respective care partner to participate in the focus group sessions.

**Survey Respondents' Characteristics**

Table 1 outlines survey respondents' characteristics. Note that total N differs between questions due to missing data, and differences in questions displayed resulting from participants' responses (e.g., if a person living with dementia did not respond for their caregiver). In total, 13 people living with dementia responded to the survey, and where applicable, responded on behalf of their caregiver. Respondents living with dementia were aged between 49 and 78 years ($M$ = 65.3 $SD$ = 9.0), and were predominantly female (69%, 9/13). In total, 62% (8/13) of people living with dementia responded that they had someone providing care for them. Respondents living with dementia reported that their caregivers ranged in age from 36 to 77 years ($M$ = 55.1, $SD$ = 15.1), and were predominantly female (63%, 5/8). The majority of these caregivers were a partner or spouse (75%, 6/8). People living with dementia were primarily diagnosed with either Alzheimer's disease (27%, 3/11) or Frontotemporal Dementia (27%, 3/11).

In addition, 64 caregivers responded to the survey and, where applicable, completed survey details on behalf of the person with dementia they were caring for. Caregivers ranged in age from 29 to 86 years ($M$ = 63.0, $SD$ = 11.2), and the majority were female (72%, 46/64). The people they cared for ranged in age from 45 to 93 years ($M$ = 78.1, $SD$ = 9.9), and over half were female (56%, 36/64). The majority of these caregivers were a spouse or partner of someone living with dementia (48%, 30/63), but there were also high numbers of adult children caring for a parent with dementia (35%, 22/63).

Table 1 – Survey Respondent's Sociodemographic Characteristics

|  | Person living with Dementia (n=13) | Caregiver (n=64) | Total (n=77) |
|---|---|---|---|
| **Person with Dementia Age*** | 65.3 (9.0) | 78.1 (9.9) | 76.0 (10.8) |
| **Person with Dementia Gender** |  |  |  |
|     Female | 9/13 (69.3) | 36/64 (56.3) | 45/77 (58.4) |
|     Male | 4/13 (30.7) | 28/64 (43.7) | 32/77 (41.6) |
| **Dementia Diagnosed** | 11/13 (84.6) | 56/64 (87.5) | 67/77 (87.0) |
| **Dementia type** |  |  |  |
|     Alzheimer's Disease | 3/11 (27.3) | 32/56 (57.1) | 35/67 (52.2) |
|     Vascular dementia | 1/11 (9.1) | 5/56 (9.9) | 6/67 (9.0) |
|     Frontotemporal Dementia | 3/11 (27.3) | 5/56 (9.9) | 8/67 (11.9) |
|     Mixed Dementia | 2/11 (18.2) | 2/56 (3.6) | 4/67 (6.0) |
|     Other | 2/11 (18.2) | 12/56 (21.4) | 14/67 (20.9) |
| **Time since diagnosis (years)*** | 4.2 (2.1) | 4.2 (2.9) | 4.2 (2.7) |
|  |  |  |  |
| **Person with Dementia has at least one Caregiver** | 8/13 (61.5) | 64/64 (100.0) | 72/77 (93.5) |
| **Person with Dementia has more than one Caregiver** | 3/8 (37.5) | 31/64 (55.4) | 34/72 (47.2) |
| **Dyad Relationship** |  |  |  |
|     Spouse/Partner | 6/8 (75.0) | 30/63 (47.6) | 36/71 (50.7) |
|     Parent/Child | 1/8 (12.5) | 23/63 (36.5) | 24/71 (33.8) |
|     Other Relative/Friend | - | 3/63 (4.8) | 3/71 (4.2) |
|     Paid Carer | 1/8 (12.5) | 7/63 (11.1) | 8/71 (11.3) |
| **Caregiver Age*** | 55.1 (15.1) | 63.0 (11.2) | 62.1 (11.9) |
| **Caregiver Gender** |  |  |  |
|     Female | 5/8 (62.5) | 46/64 (71.9) | 51/72 (70.8) |
|     Male | 3/8 (37.5) | 18/64 (29.1) | 21/72 (29.1) |

Note: n (%), *mean (SD). Total N differs between questions due to missing data, and differences in questions displayed resulting from participants' responses (e.g., if a person living with dementia did not respond for their caregiver)

**Survey Respondents' Use of Music and Technology**

With regards to music use, 55% (42/76) of people living with dementia (including those whose caregiver responded) and 75% (47/63) of caregivers used music 'often' or 'very often' in their daily lives. Further, the majority of participants (53%, 39/74 of people living with dementia; 60%, 38/63 of caregivers) felt that music was 'definitely' an important or meaningful part of their lives in the past. Survey respondents accessed music via streaming services (23%, 15/65 people living with dementia; 35%, 20/57 caregivers), CDs (35%, 23/65 people living with dementia; 47%, 27/57 caregivers) and radio (32%, 21/65 people living with dementia; 40%, 23/57 caregivers). When asked about dementia-specific applications, just 15% of total respondents (11/75) had previously used an app specific for dementia, and 62% (47/76) were willing to use a music app to support their daily activities.

Table 2 – Survey Respondent's Use of Music and Technology

|  | Person living with dementia (n=13) | Caregiver (n=64) | |
| --- | --- | --- | --- |
| **Respondent answering for** | **Themselves** | **The person they care for** | **Themselves** |
| **Current Use of Music** | | | |
|   Never | 1/13 (7.7) | 8/63 (12.7) | 4/63 (6.3) |
|   Rarely | - | 10/63 (15.9) | 7/63 (11.1) |
|   Sometimes | 3/13 (23.1) | 12/63 (19.0) | 5/63 (7.9) |
|   Often | 3/13 (23.1) | 12/63 (19.0) | 19/63 (30.2) |
|   Very Often | 6/13 (46.2) | 21/63 (33.3) | 28/63 (44.4) |
| **Music was important in the past** | | | |
|   Not at all | 2/13 (15.4) | 4/61 (6.6) | 5/63 (7.9) |
|   Somewhat | 4/13 (30.8) | 20/61 (32.8) | 20/63 (31.7) |
|   Definitely | 7/13 (53.8) | 32/61 (52.4) | 38/63 (60.3) |
|   Don't know | - | 5/61 (8.2) | - |
| **How currently listen to music** | | | |
|   CDs | 2/12 (16.7) | 21/53 (39.6) | 27/57 (47.4) |
|   Radio | 5/12 (41.7) | 16/53 (30.2) | 23/57 (40.4) |
|   Music streaming (e.g Spotify) | 5/12 (41.7) | 10/53 (18.9) | 20/57 (35.1) |
|   Video streaming (e.g. YouTube) | 5/12 (41.7) | 6/53 (11.3) | 19/57 (33.3) |
|   Digital music library (e.g. iTunes library) | 1/12 (8.3) | 9/53 (17.0) | 13/57 (22.8) |
|   TV | 2/12 (16.7) | 4/53 (7.5) | 4/57 (7.0) |
|   Wireless Speaker System | - | 2/53 (3.8) | 4/57 (7.0) |
|   Vinyl | 1/12 (8.3) | 4/53 (7.5) | 2/57 (3.5) |
|   Cassettes | 1/12 (8.3) | 2/53 (3.7) | - |
|   Sound Therapy App | 1/12 (8.3) | - | - |
|   Dementia-specific Music Player | - | 1/53 (1.9) | - |
| **Own a smartphone currently** | 11/13 (84.6) | - | 60/64 (93.8) |
| **Own a tablet currently** | 9/13 (69.2) | - | 47/64 (73.4) |
| **Experienced in use of Smartphone Apps[1]** | 7/12 (58.3) | - | 52/64 (81.3) |
| **Experienced in use of Music Apps[1]** | 5/12 (41.7) | - | 36/64 (56.3) |
| **Experienced in use of Dementia Apps[1]** | 2/12 (16.7) | - | 9/63 (14.3) |
| **Experienced in use of Sound Devices[1]** | 6/12 (50.0) | - | 32/62 (51.6) |
| **'Very' open to using a Mobile app (or a new app) as part of daily activities[2]** | 8/13 (61.5) | - | 39/63 (61.9) |
| **Ways that a music-based smartphone app might be able to support daily activities:** | | | |
|   Calming, relaxation | 5/12 (41.7) | - | 11/57 (19.3) |
|   For entertainment, enrichment, and reminiscence | 3/12 (25.0) | - | 7/57 (12.3) |
|   Improve/maintain mood | - | - | 3/57 (5.3) |
| Multiple uses | 1/12 (8.3) | - | 2/57 (3.5) |
| General benefits | - | - | 6/57 (10.5) |
| New applications must be different from other music streaming to be useful | 1/12 (8.3) | - | 14/57 (24.6) |
| Not sure/none | 2/12 (16.7) | - | 14/57 (24.6) |

Note: [1] The answers to this question initially were 'no experience', 'some experience', 'quite experienced', 'very experienced' and 'expert user'; for this table, we considered an 'experienced user' those who reported the last three options. [2] Respondents answered the question 'How open would you be to using a Mobile App (or new App) as part of your daily activities?', with either 'not open to the idea at all', 'a little open to the idea', 'very open to the idea', or 'undecided'.

When asked "What ways, if any, do you think a music-based mobile app might be able to support you in your daily activities?" people living with dementia reported an app could help with supporting calming and relaxation (41.7%, 5/12), providing entertainment or enrichment (25%, 3/12), as well as a combination of these. Caregivers responded with similar benefits, with calming and relaxation (19.3%, 11/57) and entertainment and enrichment (12.3%, 7/57) as common responses; however, 24.6% (14/57) of caregivers were also not certain about the benefits, and suggested that a new application would have to be distinct from other music technologies to be useful to support daily activities (24.6%, 14/57).

**Focus Groups**

**Focus Group Participant Characteristics**

Of those survey respondents willing to participate further in the study, 18 were able to participate in the focus groups. Focus group participants included 9 caregivers of people with dementia (5 partners, 4 adult children), 3 dyads (3 people living with dementia and 3 caregivers participating together), and 3 people currently living with dementia who attended independently. Participants ranged in age from 50 to 86 years ($M$ = 72.5 years, people living with dementia; $M$ = 70 years, caregivers); 13 participants identified as female and 5 identified as male. Table 3 provides a summary of participants.

Table 3. Focus group participants.

| Group | Pseudonym | Role | Relationship to person with dementia | Dementia Type | Music technologies discussed |
|---|---|---|---|---|---|
| 1 | Anita * | Caregiver | Partner | Alzheimer's | CDs, music streaming (with dedicated music device), records, iPad (with music from CDs) |
| 2 | Betty | Caregiver | Child | Memory problems | Radio, music streaming (on smartphone) |
| | Caroline | Caregiver | Partner | Mixed | CDs |
| | David | Caregiver | Partner | Alzheimer's | CDs |
| 3 | Elizabeth | Caregiver | Child | Alzheimer's | MP3 player |
| | Fred | Caregiver | Partner | Alzheimer's | Music streaming (on smartphone, tablet), wireless speaker system |
| | Geraldine | Caregiver | Partner of Grace | Primary progressive aphasia | Wireless speaker system, MP3 player, video streaming, digital music library, music streaming |
| | Grace | Person with dementia | Partner of Geraldine | Primary progressive aphasia | |
| 4 | Harriet | Caregiver | Child | Alzheimer's | Records, music streaming (on smartphone) |
| | Ingrid** | Caregiver | Child | Alzheimer's | Radio, CDs, cassettes, music streaming (on smartphone, tablet, to Bluetooth speaker) |
| | Ian** | Caregiver | Partner | Alzheimer's | Radio, CDs |
| 5 | James | Caregiver | Partner of Julia | Alzheimer's | CDs, radio, music streaming (to smartphone), video streaming, piano apps |
| | Julia | Person with dementia | Partner of James | Alzheimer's | |
| | Karen | Caregiver | Partner of Kevin | Lewy Body | |
| | Kevin | Person with dementia | Partner of Karen | Lewy Body | Radio, music streaming |
| | Lucy | Person with dementia | Attended independently | Mixed | CDs, MP3 player |
| | Mary | Person with dementia | Attended independently | Mixed | Radio, CDs |
| | Naomi | Person with dementia | Attended independently | Fronto-temporal | Radio, records |

Note. * Focus group 1 was conducted with one participant as two others who were scheduled to join withdrew due to unforeseen circumstances. ** Caregivers Ian and Ingrid are father and daughter, both caring for the same person – Ian's partner/Ingrid's mother, who did not participate in the study.

**Benefits of Music Technologies: How Caregivers and People living with Dementia use music as part of their everyday care practices**

Caregivers and people living with dementia described four benefits of using music technologies in everyday care: 1) to regulate mood and provide joy, 2) to facilitate music-based social activities, 3) to provide continuity before and after a clinical diagnosis, and 4) to make caregiving easier.

**Using music for regulating mood and providing joy**

Caregivers said they often used music and music technologies as a tool for regulating the mood of their loved ones in their everyday lives and in care practices. Caregivers believed that music was valuable for its ability to help manage the behavioural and psychological symptoms of dementia and provided joy for both caregivers and people living with dementia. Caregivers chose to listen to music out loud through shared technologies (e.g., speakers) because they perceived these potential benefits for their loved ones.

Harriet, for example, found that listening to vinyl records could be valuable in managing her mother's mood:

*'I would say [music] lifts her mood…she does get a bit down…gets a bit paranoid sometimes, but I just think music has a calming effect and it does definitely lift her mood.'*

For Fred and his partner, music had a relaxing effect even when *'choosing different playlists* [that are] *not specific for calming'*. Ingrid, who cared for her mother during the day, noted the benefits of music listening as part of a routine, and at difficult times of day, notably "sundowning":

*'And if the light is changing, and I know that there's still two to three hours before Papa arrives to collect her….I say: "shall I put some music on?" And she said: "Oh yes, please." And I find some classical works extremely well, and anything that's calming...gentle jazz…'*

As well as specific emotional impacts of music, caregivers and people with dementia discussed the benefits of using music to manage behavioural symptoms of dementia. For example, Ian, also speaking about Ingrid's mother, stated that:

*'… [she is] a bit prickly at the moment…and then occasionally she will lash out at those closest to her, so I use music as a pacifier.'*

In terms of the technologies used to access music, their portability provided benefits for enabling music to be used when and where it was needed, as part of emergent care practices for both the person living with dementia and their caregivers. Lucy, a person living with dementia, said that she found it useful to keep a small MP3 player *'handy'* or close by at night to help her manage a particularly troubling symptom of dementia, hallucinations that occurred at night. Lucy found that using her MP3 player to listen to music helped her to manage these hallucinations.

Furthermore, the portability of newer music technologies like MP3 players, or streaming services such as Spotify, made music an accessible means of emotion regulation while outside of the home. Anita, caring for her partner, discussed the benefits of streaming music from a mobile device:

*'It's also then something that I can take with us, you know…there are occasions when we have to go to appointments and things like that and if there is waiting time then you know I can use it there…if I feel him getting a bit agitated or anything'.*

While music listening can have a range of practical benefits for people with dementia and their caregivers, participants also noted the fun and enjoyment that could arise from actively participating in and interacting with musical activities. Active forms of engagement like singing, playing instruments, and moving to music were discussed. In multiple examples, participants with dementia and caregivers spoke of the joy they experienced when singing to music. For example, Naomi, a person living with dementia, stated:

*'Now because I'm having struggle speaking, but I have the joy of being able to sing…it's so much easier to sing now than actually talk some days.'*

Other enjoyable activities included moving and dancing to music. Geraldine, a caregiver, spoke of how she and her partner could enjoy music and dancing together by streaming music with wireless speakers in their home:

*'We have a playlist that we both like and every now and then…one of us might go to the other person's room wherever they're cleaning you know and have a dance because it's…particularly memorable for us.'*

These examples suggest benefits that music activities beyond listening can offer for people with dementia and their caregivers. Participants described how everyday and spontaneous activities, such as singing and dancing to music, can be supported using technologies that are available within the home (e.g. wireless speakers).

**Technologies can facilitate music-based social activities in care**

Music and music technologies were reported to encourage social interaction. Participants shared examples that showed how participating in organised music activities facilitated social bonding for those still living at home. These activities were supported by technology that provided access to music resources beyond the setting of the organised activity.

In the home care context, Geraldine, a caregiver, described attending a social musical activity for people with dementia and their caregivers:

*'we had another session which was a karaoke sing along for people and it was…through a care group so it wasn't necessarily all for people with dementia, but it was for people caring for other people…but any kind of interaction I think brings so much more to a person's life, if they are involved in doing something while the music is playing.'*

These musical activities were sometimes extended with technologies used in the home. For example, James, who cared for his partner at home, shared how they used a music app to assist with singing in their local choir:

*'I use certain apps to do the choir sections…when you sing in a choir, it's very hard to get your pitch right. So, you use your app, that's why I've got a keyboard to try and play the parts of the choir. So, you download, basically using YouTube and the facilities there they give you choir sections which break up the score into alto, soprano and so on.'*

In other words, James and his wife used resources available through YouTube to practice their parts for the choir.

Music technologies were important, as they not only enabled participants to share music with others – for example, by listening to music together – but also made it possible to experience music and a sense of companionship independently. Ingrid, who cares for her mother with dementia and also hosts a radio show, highlighted the role that music technologies, such as radio, can play in fulfilling social needs.

*'So, radio again…it's very personal… literally there is somebody in the room there and you don't have to recognise them. You can recognise the voice, for example, you can hear, say, certain voices are absolutely unmistakable when you hear them on radio'.*

**Music provides continuity before and after diagnosis and encourages reminiscence**

One of the key benefits caregivers shared was that music technologies enabled continuity in music listening throughout different stages of their experience with dementia. Caregivers and people living with dementia described their existing relationships with music and explained that listening to music provided a link between experiences before and after dementia diagnosis.

Participants living with dementia primarily preferred listening to older styles of music from their youth, yet caregivers also noted the potential benefits of other styles. Music from earlier decades was often familiar, and despite experiencing difficulties with memory, people with dementia still remembered songs from their youth and enjoyed singing along to them. Kevin, who has Lewy Body dementia, noted:

*'With the Lewy Body, it doesn't affect your long-term memory really, it's the short-term memory. So, a lot of the old songs, you can remember. It just comes out and you sing'.*

While there was substantial emphasis on jazz and classical music as genres people with dementia *'used to like listening to'* (Caroline), participants also noted that music taste was not fixed, and that other genres could provide benefits. Elizabeth, who cared for her mother, noted:

*'I had this…idea that, you know, mum loves Ella Fitzgerald and all those old female singers and Frank Sinatra…and I sort of had this, right it's gotta be all those oldies or Peter Allen 'cause she loves him and those songs from the 70s. But I had some success with… like one of my kids would play Taylor Swift or something…and that sort of calmed her'.*

Music collections were predominantly gathered in the years prior to dementia diagnosis, and music from individuals' formative years was prominent. The technologies used were diverse across participants, often reflecting participants' music use prior to diagnosis. For example, Harriet noted that her mother had *'a big record collection'*, while Anita's husband had *'CDs from…singers he would have grown up listening to'*. Radio was popular with both people living with dementia and caregivers, particularly stations that played classical music. For example, Lucy shared that radio *'kept [her] sane'* during the COVID-19 pandemic, while caregivers, such as James and Ingrid, discussed turning on the radio regularly as it was a familiar technology:

*'So music in radio is very, very powerful because that's also a medium, in actual fact it was the medium, apart from cinema, for my parents' generation growing up'.* (Ingrid).

In addition, Fred used streaming services, with *'playlists that [were] put together some years ago before the dementia set in'*.

Notably, despite the positive effects of music listening, participants also recalled negative experiences. In particular, music could be associated with memories that were not always welcome for the person living with dementia. After trialling an MP3 player and headphones with her mother, Elizabeth noted:

*'For my mum it [music] triggered a lot of memories and she wanted to shut them down, and music upset her very much'.*

For many caregivers, playing music was used as an important way to facilitate reminiscence and, as Ian described, *'a different form of bringing back reality'* for their loved ones. For example, Ingrid (speaking about her mother, Ian's spouse) had a sense that music was special for her mother, sharing that they often play the radio for her:

*'Sometimes she doesn't recognise me but recognises my voice. So, I do a jazz programme on community radio every 3-4 weeks…and dad…will have the radio on and she will recognise my voice, and…[she] will recognise the music and the pieces, particularly if they are ones that I was familiar with growing up…So…there will be like a little trigger there'.*

Elizabeth discussed a situation when she visited her mother in residential aged care, and other residents began sharing stories with her about their memories associated with music:

*'I actually had little sort of meetings in the living room of the nursing home, and it turned into kind of a storytelling session… other residents would come up to me and…tell me their musical memories of gigs because…I told them all that I was putting together a playlist for my mum. So, what I loved was all the storytelling of the gigs from many years ago…and that was so beautiful…for people to share that because they were all at varying stages in their dementia journey'.*

In this example, using technology to facilitate music listening – that is, "putting together a playlist" – triggered a social storytelling session focused on reminiscing about past music experiences. Participants highlighted how streaming services and MP3 players can be used to create tailored playlists that foster reminiscence for people living with dementia, but these playlists need to be used carefully, given that music can also provoke painful memories, as highlighted by Elizabeth. While the link between music and reminiscence is well-known, these examples show that family caregivers are informally using music technologies to connect with their loved ones, including being mindful of both the positive link with past experiences that music can provoke and the risk of triggering difficult memories when listening to music.

**Music technologies help make caregiving easier**

As well as using music to directly support their loved ones, caregivers also used music technologies to address their own needs. By providing entertainment and relaxation, music freed up time for other activities, and helped to make caring easier for both participants and others providing care for their loved ones. In both situations, new music technologies made it easier for caregivers to use music strategically in their caregiving.

When used to provide an enjoyable distraction for the person living with dementia, music freed up caregivers' time to attend to other activities. Participants explained that newer technologies (e.g. streaming services), had advantages over older modalities (e.g. records) in this regard, because they could be left to play for longer. As Anita explained:

*'We have lots of old records…but I find that if I play the music there, I've no sooner got it on than it seems to stop…[it] interrupts another activity that I might be involved in so I find the newer technologies leave me freer to do other things here at home for a longer time'.*

Music was utilised not just by the primary caregiver, but also by others in the care network. Fred had set up music technologies in the home prior to his wife's diagnosis. This allowed visiting caregivers to use these technologies to play music when caring for Fred's wife:

'There's a carer in each day…and if I'm out they can still tap into [the wifi] and play music that's on their [phone].….That's another big win now, exploring with that so they can get music going for [my wife] and either their own music, their own playlists, individual songs, etcetera.'

Finally, caregivers described using music technology to entertain their loved one. David mentioned putting on music in the background so that his partner was not sitting in silence. Anita said that she frequently plays CDs at home as a way to fill spare time:

'I just use it in those times, when there's really nothing else [to do] apart from sleep and just sitting staring.'

**Barriers for Music Technology in Care**

Despite the potential advantages of using music listening technologies in care, several issues with technology use were raised by participants. Barriers to use centred around a difficulty for caregivers and people living with dementia to stay up to date with constantly evolving technologies, and issues with confidence and ability to use digital technologies.

**Difficulty staying up to date with evolving technology**

Participants described difficulties staying up to date with evolving technology in terms of smartphone applications and other music technologies. Even if familiar, these sometimes became more difficult for people living with dementia and caregivers to use over time.

As a self-described older caregiver, James said he experienced difficulties with the planned obsolescence common to new digital technologies:

'To try to adapt as you're older to the phones, and they're designed to be out of date within six years and you gotta find a new one and it all changes, I just find it…very difficult to negotiate.'

Caregivers faced similar barriers with smartphone applications. Geraldine mentioned an instance when she and her partner with dementia stopped using an app that they had previously found useful. The app was practical until they changed from one smartphone brand to another where the app displayed differently, and they encountered errors:

'[Grace] had an iPhone at the time and then changed over to Samsung, and it was an Apple App. The Samsung version is not nearly as user friendly…the Apple one was just visually clearer…seemed to be easier to upload [photos]…this one…got stuck…it wouldn't let me [upload]…So Grace doesn't use that so much now.'

When discussing other music technologies, Harriet said her mum still relies on records, while David, caring for his wife, stated that he still uses CDs and doesn't know how to use newer forms of music technology:

*'I would prefer CDs I think because they're controllable…and there are different ones. Presumably you can do that all digitally but I don't really know how to do that.'*

Similarly, Lucy, a person living with dementia, often listened to the radio, but believed that other digital technologies might make music listening easier if she could learn how to use them. Lucy did use an MP3 player, but faced difficulties downloading music onto the device:

*'It's like, I know music, listening to it in those little…iPods…because it's handy in the night if I feel like the traumas are coming up or some hallucinations. Immediately put that [on], it will at least calm me down. But I don't have the know-how, how to do that…the downloading…I'm not very good at.'*

**Participants with dementia expressed low self-efficacy with technology**

While older technologies tended to be more familiar, people with dementia still faced challenges with using some technologies, and expressed low self-efficacy with technology use. For example, Harriet found that her mother often had trouble putting on a vinyl record:

*'I think she panics. She panics with anything with any buttons. If she calms down enough, she'll do it. But if she starts overthinking it, then she just panics and can't do it.'*

Mary, a person living with dementia, stated that it was common for people with dementia to feel overwhelmed when talking about smartphone technologies:

*'As you talk 'app', nine tenths of the people I know with dementia are going to back away because that's beyond them.'*

As a result of these challenges, several caregivers discussed how important it was that applications and smartphone-based technologies were easy to use. In one example, Harriet stated:

*'I'm just all about functionality: easiness. Basically, the easier the better. The more basic, the better'.*

The accessibility of smartphone technology was also considered at length, with participants discussing hearing issues (Betty) and vision impairment. For example, James stated: *'[a] bigger screen is definitely helpful rather than small screens'*. There were also physical issues with using mobile technology. Ingrid noted: *'if I'm typing [on the phone]… I really have to be very careful, how I type, and…that's just me [a younger person] so for somebody who no longer has agility, let alone mobility with fingers that could be an issue'*.

In addition, some participants perceived smartphone applications as difficult to use for people living with dementia. For example, Betty was concerned that setting up a music app for her mother with dementia would require *'knowledge, information or training for the person setting it up, which probably may well be the carer.'*

## Discussion

**Principal Findings**

Employing a mixed-methods design, this study aimed to understand how people living with dementia and their family caregivers use music and music technologies in their everyday caring, as well as the challenges they experience using music technologies. In our survey, we found that the majority of participants living with dementia and their caregivers use music often in their daily lives, with music an important or meaningful part of their lives in the past. Consistent with prior work, participants reported a range of methods and devices for accessing music, with high rates of use for music and video streaming services, as well as use of more traditional music technologies such as CDs and radio [40]. Participants were interested in new ways of using music to support their care, with a willingness to try new music-based mobile applications, and considered a range of potential benefits such an app could provide. These findings, alongside previous work supporting the important role music can play in the everyday lives of people living with dementia [24], suggest that new music-technology solutions may provide valuable benefits for this group.

With further focus groups, we found several perceived benefits of using music technologies, alongside challenges that caregivers and people living with dementia face in using music and technology in these contexts. We found that people living with dementia and their caregivers use music for mood regulation and providing joy, encouraging social connection, to make caregiving easier, and to provide continuity and promote reminiscence. This highlights that the everyday ways people living with dementia and their caregivers use music in care align with benefits identified in research evaluating music interventions with people living with dementia [11,23,25,28,31]. We also identified barriers to using music technologies, predominantly due to participants with dementia having low self-efficacy with technology use, and challenges for both caregivers and people living with dementia in staying up to date with evolving technology. Here, we consider how future music technologies can be designed to capitalise on the existing uses of music technology among caregivers of people living with dementia, and to empower caregivers to use music strategically and therapeutically at home [7].

People living with dementia and caregivers predominantly used music for mood regulation, such as relaxation, to lift mood, and to provide joy. This is consistent with prior research finding that caregivers and people living with dementia often use music for specific functions, such as relaxation or calming [23]. For example, Ingrid listened to music with her mother to manage behaviours such as late day confusion and restlessness. In addition, participants spoke fondly of the capacity for music to simply provide joy, echoing previous work where music and sound technologies have been shown to enrich the lives of people living with dementia, and provide enjoyable moments at home [29]. This highlights the need for music technologies and interventions to consider the enjoyable, recreational aspect of music as well as its potential therapeutic benefits [11]. Importantly, digital technologies provide an avenue for these benefits to be translated to situations outside of the home, with Anita and Lucy suggesting that music could be a convenient, portable mood regulation tool.

The role of music and music technologies for supporting social connection, both within a care dyad and with others, was evidenced in our study, and aligns with prior research [25,29,31–33]. Participating in musical activities outside of the home fostered social connections for people living with dementia, as did musical activities within residential aged care. Technology was important to these interactions. For example, James discussed how mobile applications supported him and his partner in practicing their choir singing.

We found that caregivers and people living with dementia described existing relationships with music, outlining some continuity between experiences before and after dementia diagnosis. Music technologies facilitated this continuity, enabling caregivers to encourage use of familiar technologies, such as CDs and vinyl records, as well as personalised playlists. Music use was often associated with reminiscence, with people living with dementia reporting that they could sing along to old favourites, even if they experienced challenges with memory or other forms of verbal communication. Consistent with the idea of the "reminiscence bump", where individuals tend to favour the music of their youth [41–43], many people living with dementia preferred older styles of music. Existing relationships with music were also connected with existing collections of records, CDs, and favourite radio stations, with our findings also illustrating the importance of radio for older adults [44,45].

Finally, consistent with previous work [25,46], music technologies were important for making caregiving easier, such as accessing music to fill time or silence. In addition, modern music technologies, such as wireless speakers, enabled music to be played for the person living with dementia not just by the primary carer, but by others in the broader care network such as in visiting caregivers, with minimal setup.

**Barriers and Design Opportunities**

While previous research has identified uses of music in residential aged care [13,23], and designed tools for music listening, sound, and music activities in dementia care [27–30,33,34], there has been limited research investigating the everyday uses of music technologies in care practices by family caregivers of people living with dementia. To date, no research has clearly identified opportunities for music technologies to play a role in the therapeutic use of music at home. Building on the current findings, ensuring ease of use of music technologies is likely crucial for engaging caregivers and people living with dementia in music use, both for its therapeutic benefits and for recreational uses of music. This is especially pertinent given that the major barriers to music technology use in everyday care included difficulties with self-efficacy around technology use, and challenges staying up to date with evolving technology. Previous research has established the importance of digital literacy and self-efficacy in technology use among older adults [47], indicating that this is an important consideration when designing technologies for people living with dementia. Therefore, if caregivers and people living with dementia are not confident utilising existing technologies to support music use, co-designing new technologies that capitalise on the everyday ways these groups use music technologies should be a priority.

Our findings suggest several design opportunities for future technologies. New technologies should be designed to align with how music is used to regulate mood, and capitalise on the ways that music technologies can be used as a portable emotion regulation tool. Music technologies can also be used to foster social interaction between people living with dementia and their caregivers, with video conferencing tools and apps designed to support virtual music activities, such as choirs and dancing. Additionally, in designing music technologies for people living with dementia and their caregivers, access to personally preferred music is crucial. Often this may be from the reminiscence bump (i.e. music from a person's youth), however music taste should not be assumed, nor limited only to older styles. These considerations for music selection and the affordances of personalised playlist creation highlight the benefits that streaming platforms might have over physical music collections, as long as they are easy for both people living with dementia and caregivers to access. One way this could be achieved is through playlists configured to play on a smart speaker with an easy-to-remember prompt that activates the desired playlist. Research has shown that voice interaction devices such as smart speakers (e.g., Amazon Alexa, Google Home) have accessibility advantages over screen-based technologies for some users [48–51]. Music technologies can also be designed to make it easy for caregivers to play recorded music, and use music in intentional ways, to relieve pressure for family caregivers [6,7]. Focusing on such factors, and ensuring that people living with dementia and their caregivers are comfortable using such technologies, is a crucial first step to supporting the use of music technologies in care.

**Conclusions**

Informal caregivers report that they need additional support and information [4], yet they are often reluctant or lack knowledge to fully utilise services [5], and many family caregivers of people living with dementia experience high levels of stress and burden [6]. As a result, caregivers and people living with dementia need accessible, everyday strategies to draw on to support their needs, and technologically supported music use is one key avenue for future research [7,9,27]. This study draws attention to the ways that these groups already use music technologies to support their needs, outside the context of organised activities or targeted music interventions. Caregivers and people living with dementia believe that music technologies may be a useful way to access music for regulating mood, providing joy, encouraging social connection, providing continuity, promoting reminiscence, and making caregiving easier. The technologies used to play music are diverse across users, and people living with dementia and their caregivers face barriers to using new digital music technologies, despite acknowledged advantages over more familiar technologies. As such, there is a clear opportunity to design technologies to capitalise on existing music use in these groups, to enable easier access to specific music activities to support people living with dementia with recreational and therapeutic music listening and music-based activities.

**Author contributions:**

The overall project was developed by FB, with contributions from TVS, JT, and JW. RC, TVS, JT, JW, and FB designed the study. RC managed focus group data collection. KM assisted RC in focus group data collection and PASS contacted participants for the focus groups. Survey data was summarised by TVS, with further refinement by DV during writing. TVS and RC transcribed focus groups. Transcripts of focus groups were coded by DV using an inductive approach. After familiarisation with the transcripts, DV generated initial codes and developed initial themes, with JW and RK reviewing and refining themes into the final set. DV, RK, and JW defined and named the final themes, further refining with the rest of the authors during writing. All authors approved and edited the final version of the manuscript.

**Conflicts of Interest:**

The authors have no conflicts of interest to disclose.